\documentclass[pre,twocolumn,floatfix, superscriptaddress,a4paper,nofootinbib]{revtex4-1}
\usepackage{amsmath,bm,graphicx}
\usepackage{amssymb}
\usepackage{amsfonts}
\usepackage{epstopdf}
\usepackage{mathrsfs}
\usepackage{color}
\usepackage{latexsym}
\usepackage{hyperref}
\usepackage{subfigure}
\usepackage{bm}
\usepackage{natbib}
\usepackage[normalem]{ulem}

\begin{document}

\title{Transport of neutral and charged nanorods across varying-section channels}

\author{Paolo Malgaretti}
\email[Corresponding Author: ]{p.malgaretti@fz-juelich.de }
\affiliation{Max-Planck-Institut f\"{u}r Intelligente Systeme, Heisenbergstr.~3, D-70569
Stuttgart, Germany}
\affiliation{IV. Institut f\"ur Theoretische Physik, Universit\"{a}t Stuttgart,
Pfaffenwaldring 57, D-70569 Stuttgart, Germany}
\affiliation{Helmholtz Institute Erlangen-N\"urnberg for Renewable Energy (IEK-11), Forschungszentrum J\"ulich, F\"urther Stra{\ss}e 248,
90429 N\"urnberg, Germany}

\author{Jens Harting}
\affiliation{Helmholtz Institute Erlangen-N\"urnberg for Renewable Energy (IEK-11), Forschungszentrum J\"ulich, F\"urther Stra{\ss}e 248,
90429 N\"urnberg, Germany}
\affiliation{Department of Chemical and Biological Engineering and Department of Physics, Friedrich-Alexander-Universit\"at Erlangen-N\"urnberg, F\"{u}rther Stra{\ss}e 248, 90429 N\"{u}rnberg, Germany
}

\begin{abstract}
We study the dynamics of neutral and charged rods embedded in varying-section channels. By means of systematic approximations, we derive the dependence of the local diffusion coefficient on both the geometry and charge of the rods. This microscopic insight allows us to provide predictions for the permeability of varying-section channels to rods with diverse lengths, aspect ratios and charge. Our analysis shows that the dynamics of charged rods is sensitive to the geometry of the channel and that their transport can be controlled by tuning both the shape of the confining walls and the charge of the rod. Interestingly, we find that the channel permeability does not depend monotonically on the charge of the rod. This opens the possibility of a novel mechanism to separate charged rods.
\end{abstract}

\maketitle

\section{Introduction}

The transport of molecules, proteins and small particles across pores,
channels and, in general, porous materials is of paramount relevance
in several biological, environmental and technological scenarios~\cite{Malgaretti_2019}.
In biology and technological applications, ions and molecules are transported across membrane pores~\cite{Calero,Roth2014,Peukert2017},
RNA is transported across the nuclear membrane~\cite{Gracheva2017,Bacchin2018,Dagdug2019} and the crowded environment of cell cytoplasm provides dynamical obstacles to the motion of organelles and vesicles~\cite{Albers}. Similarly, in environmental sciences, pollutants and plant nutrients spread across the porous matrix of rocks and soil. 
In this perspective, substantial effort has been dedicated to the synthesis~\cite{Chao2017,Chen2019} and characterization~\cite{Yang2019} of anisotropic nano-colloids and stiff filaments~\cite{Fakhri2010,nanocolloidsBook}.
Finally, in recent years the development of experimental techniques such as DNA~\cite{bonthuis2008conformation,Sakaue_2018} and protein~\cite{Chinappi_2018} 
sequencing, resistive-pulse sensing techniques~\cite{Saleh2003,Ito2004,Heins2005,Arjmandi2012} and  chromatography~\cite{MARTINEZCRISTANCHO201629,Ghosh2002,Kaspereit2014} have been developed by exploiting transport across pores. 
%\jh{Another example could be the chromatography of nanoparticles as in CRC1411.}
%\pa{for me it is fine. Do you have some references at hand?}
%\jh{I started here: https://www.sciencedirect.com/science/article/pii/S0032591018306107
%The paper has a section on chromatogrpahy in the intro, check for references with Peukert or Kaspereit or Seidel-Morgenstern}

On the theoretical side, tackling these problems is typically complicated since it requires detailed numerical simulations that take into account
the interactions between the suspended object and the confining walls.
Recently, several groups have contributed developing the so-called Fick-Jacobs
approximation\cite{Zwanzig,Reguera2001,Kalinay2008,martens2011entropic,Dagdug2013,Malgaretti2013}. This scheme relies on the assumption that when the net longitudinal velocity is slow enough, the transported object will explore the directions
transverse to the motion with a probability distribution that is very
close to the equilibrium one. Thanks to this approximation, the geometry
of the pore as well as the interaction with the walls can be accounted
for by the local equilibrium free energy. This approach
has been shown to be applicable to the transport of colloids~\cite{Reguera2006,Reguera2012,Marconi2015,Malgaretti2016_entropy,Puertas2018},
ions~\cite{Malgaretti2014,Malgaretti_macromolecules,Malgaretti2015,Chinappi2018,Malgaretti2019_JCP}, and polymers~\cite{Bianco2016,Malgaretti2019,Ceccarelli2019}  just to mention a few among others (see recent review articles~\cite{BuradaReview,Malgaretti2013} for a more comprehensive list). 

\begin{figure}
\includegraphics[scale=0.5]{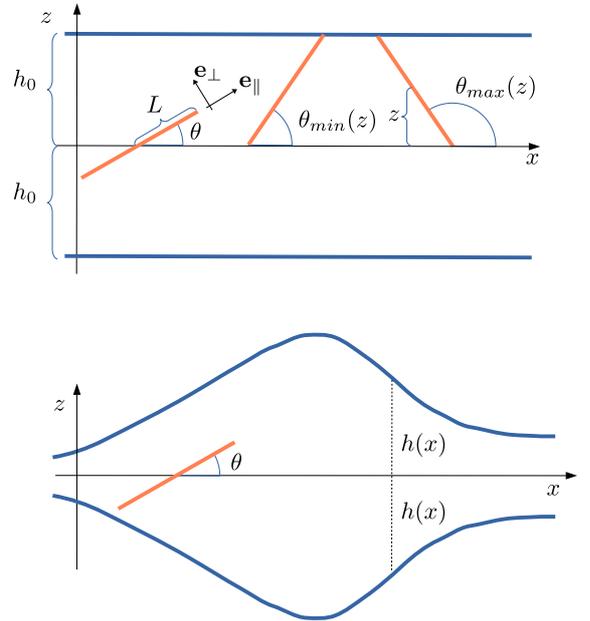}
\caption{Top: scheme of a rod (orange) with major axis of length $2L$ tilted by an angle $\theta$ and embedded in a plane channel with half section $h_0$. The minimum, $\theta_\text{min}(z)$ and maximum, $\theta_\text{max}(z)$ angles for a given position of the center of mass are reported. Bottom: a rod in a varying-section channel with half-section $h(x)$.}
\label{fig:1}
\end{figure}

In this article we study the dynamics of charged rods confined in varying section channels. In particular, we propose an extension of the Fick-Jacobs approximation that allows us to account for both the geometry of the rod as well as for its net charge. From this perspective, our work goes beyond a recent contribution (see Ref.~\cite{FJ-rods}) in which an approximated equation governing the transport of uncharged rods across varying section channels has been proposed.
First, we test our approach against known experimental results~\cite{Hanggi2019} for neutrally charged rods.
Our model retrieves both the local dependence of the diffusion coefficient as well as the dependence of the Mean First Passage Time (MFPT) profile. 
Second, we use our model to predict the dependence of the diffusion coefficient on the geometry of both the channel and the rod as well as on the charge of the rod. 
For neutrally charged rods, in agreement with Ref.~\cite{Hanggi2019}, we find that the local diffusion coefficient maximizes at the channel bottleneck, i.e. where the rod is mainly oriented along the channel axes. Interestingly, the modulation of the local diffusion coefficient is maximized when the length of the rod major axis is comparable to the channel average section and it becomes vanishing small when these two length scales differ. %Moreover, our data show that the net diffusion in a varying section channel can be enhanced as compared to the case of a plane channel.
Finally, our data show that such a dependence can be even amplified by the electrostatic interaction between the rod and the channel walls. 

\section{Model}

We characterize the transport of rigid rods across a $2D$ varying-section
channel whose half-section is given by 
\begin{equation}
h(x)=h_{0}\left(1-h_{1}\cos\left(\frac{2\pi}{L_{0}}x\right)\right).
\end{equation}
where $h_0$ is the average section, $h_1$ is the amplitude of the section modulation and $L_0$ is the period of the channel. 
Accordingly, the $2D$-Smoluchowski equation reads
\begin{equation}
\partial_{t}\rho(x,z,\theta,t)=\partial_{x}J_{x}+\partial_{z}J_{z}+\partial_{\theta}J_{\theta}\,,
\label{eq:smol-1}
\end{equation}
with 
\begin{eqnarray}
J_{x} & = & \mathbf{e}_{x}\cdot\mathbf{D}(\theta)\cdot\left[\bar{\nabla}\rho+\rho\beta\bar{\nabla}W\right]\label{eq:Jx},\\
J_{z} & = & \mathbf{e}_{z}\cdot\mathbf{D}(\theta)\cdot\left[\bar{\nabla}\rho+\rho\beta\bar{\nabla}W\right],\label{eq:fluxes}\\
J_{\theta} & = & D_{\theta}\partial_{\theta}\rho\,,
\end{eqnarray}
where $\rho$ is the probability distribution function, $\beta^{-1}=k_BT$ is the inverse thermal energy, $k_B$ is the Boltzmann constant, $T$ is the absolute temperature. Moreover, $\mathbf{D}(\theta)$ is the translational diffusion matrix that accounts
for the fact that the diffusion along the major axis of the rod differs
from that perpendicular to it, $D_{\theta}$ is the rotational diffusion
coefficient and $W$ accounts for the both, the conservative forces acting on the rod and the confinement. 
In the following, we focus on the
case in which the local radius of curvature is much larger than the
length of the rod. In such a regime the channel walls can be approximated
as locally parallel to the longitudinal axis. Accordingly we have
that the potential $W$ reads
\begin{equation}
W(x,z,\theta)\!=\!\begin{cases}
\phi(x,z,\theta)-fx & \!\!|z|<h(x)\,\,\&\,\,\,\theta_{m}\!<\!\theta\!<\!\theta_{M}\\
\infty & \text{else}
\end{cases}
\end{equation}
with 
\begin{eqnarray}
\theta_{m} = & \begin{cases}
\frac{\pi}{2}-\arccos\left(\frac{h(x)-z}{L}\right)\!\! & \!h(x)-L\leq|z|\leq h(x)\\
0\!\! & |z|<h(x)-L
\end{cases}\label{eq:theta_m}\\
\theta_{M} = & \begin{cases}
\frac{\pi}{2}+\arccos\left(\frac{h(x)-z}{L}\right)\!\! & \!h(x)-L\leq|z|\leq h(x)\\
\pi\!\! & |z|<h(x)-L
\end{cases}\label{eq:theta_M}
\end{eqnarray}
where $\phi(x,z,\theta)$ accounts for the equilibrium conservative forces and $f$ for the longitudinal force responsible for the transport along the channel. In Eqs.~\eqref{eq:theta_m},~\eqref{eq:theta_M} we assume that the
aspect ratio between the long axis, $2L$, and the minor axis, $2l$, of the rod 
is such that $L\gg l$ and hence we can disregard the geometric corrections
to Eqs.~\eqref{eq:theta_m},~\eqref{eq:theta_M} due to the finiteness of $l$.

We further assume that translation along the channel axis
is slow enough such that $\rho$ retains its equilibrium profiles along
$z$ and $\theta$, which implies the absence of fluxes in $z$ and $\theta$, i.e.% we assume% that we have
\begin{eqnarray}
J_{z} & = & 0\,,\label{eq:Jz}\\
J_{\theta} & = & 0\label{eq:Jtheta}\,.
\end{eqnarray}
Accordingly, we perform the standard Fick-Jacobs approximation~\cite{Zwanzig,Reguera2001,Kalinay2008,martens2011entropic,Dagdug2013} 
\begin{align}
\rho(x,z,\theta,t) & =p(x,t)\frac{e^{-\beta W(x,z,\theta)}}{e^{-\beta A(x)}}\,,
\label{eq:FJ0}
\end{align}
where 
\begin{align}
\beta A(x) & =-\ln\left[\frac{1}{2 h_0 \pi}\int\limits_{0}^{\pi}\int\limits_{-\infty}^{\infty}e^{-\beta W(x,z,\theta)}dzd\theta\right],
\end{align}
is the local equilibrium free energy~\cite{Zwanzig,Reguera2001}. 
%Integrating Eq.~\eqref{eq:smol-1} along $z$ and $\theta$ and using Eqs.~\eqref{eq:Jz},\eqref{eq:Jtheta} we obtain
%\begin{align}
%J_{x} & =\mathbf{e}_{x}\cdot\mathbf{D}(\theta)\cdot\mathbf{e}_{x}\left[\partial_{x}\rho+\rho\beta\partial_{x}W\right]\,.
%\end{align}
%i.e., the extension of the Fick-Jacobs approximation to the case of rigid rods embedded in varying-section channels. 
\begin{figure*}
\includegraphics[scale=0.4]{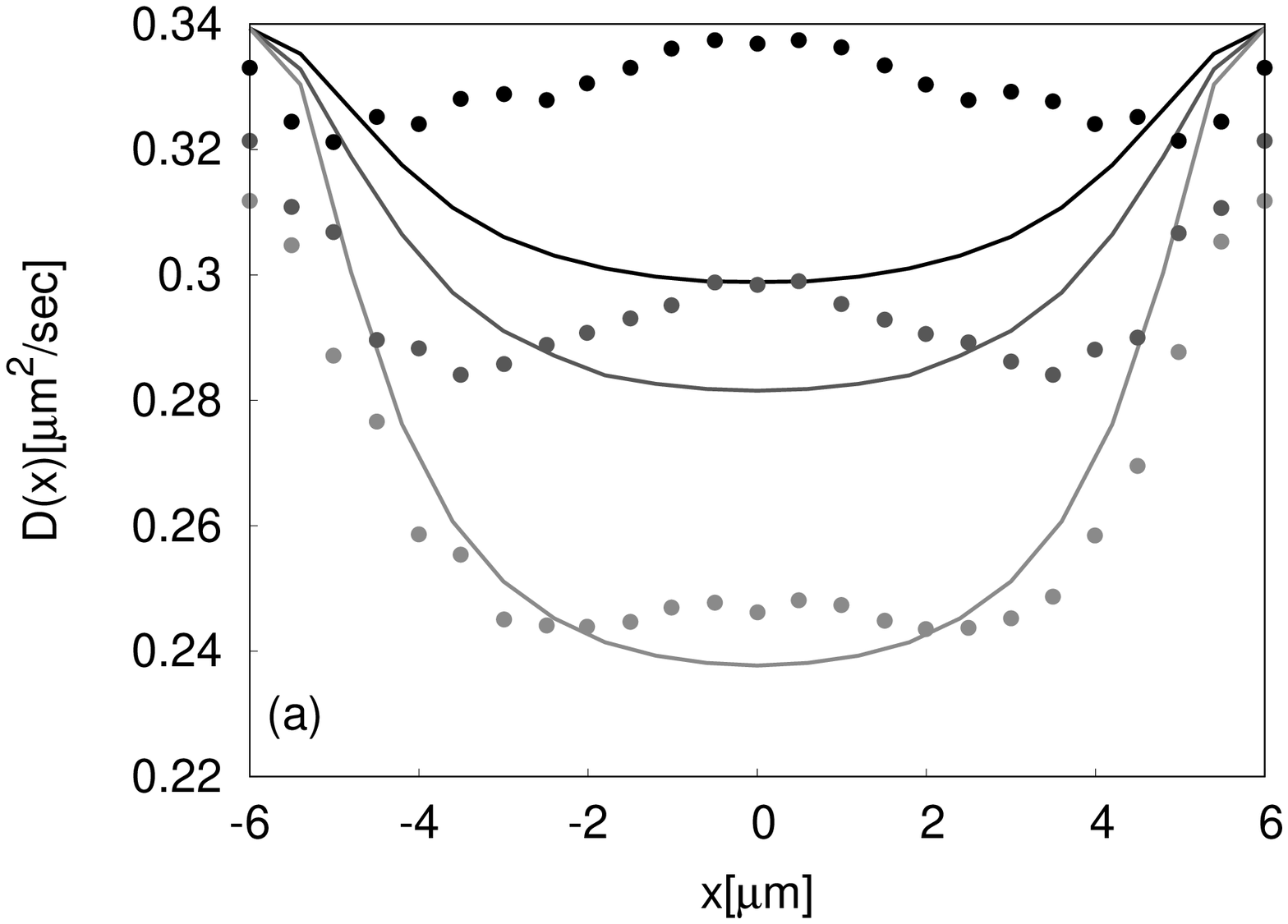}
\includegraphics[scale=0.4]{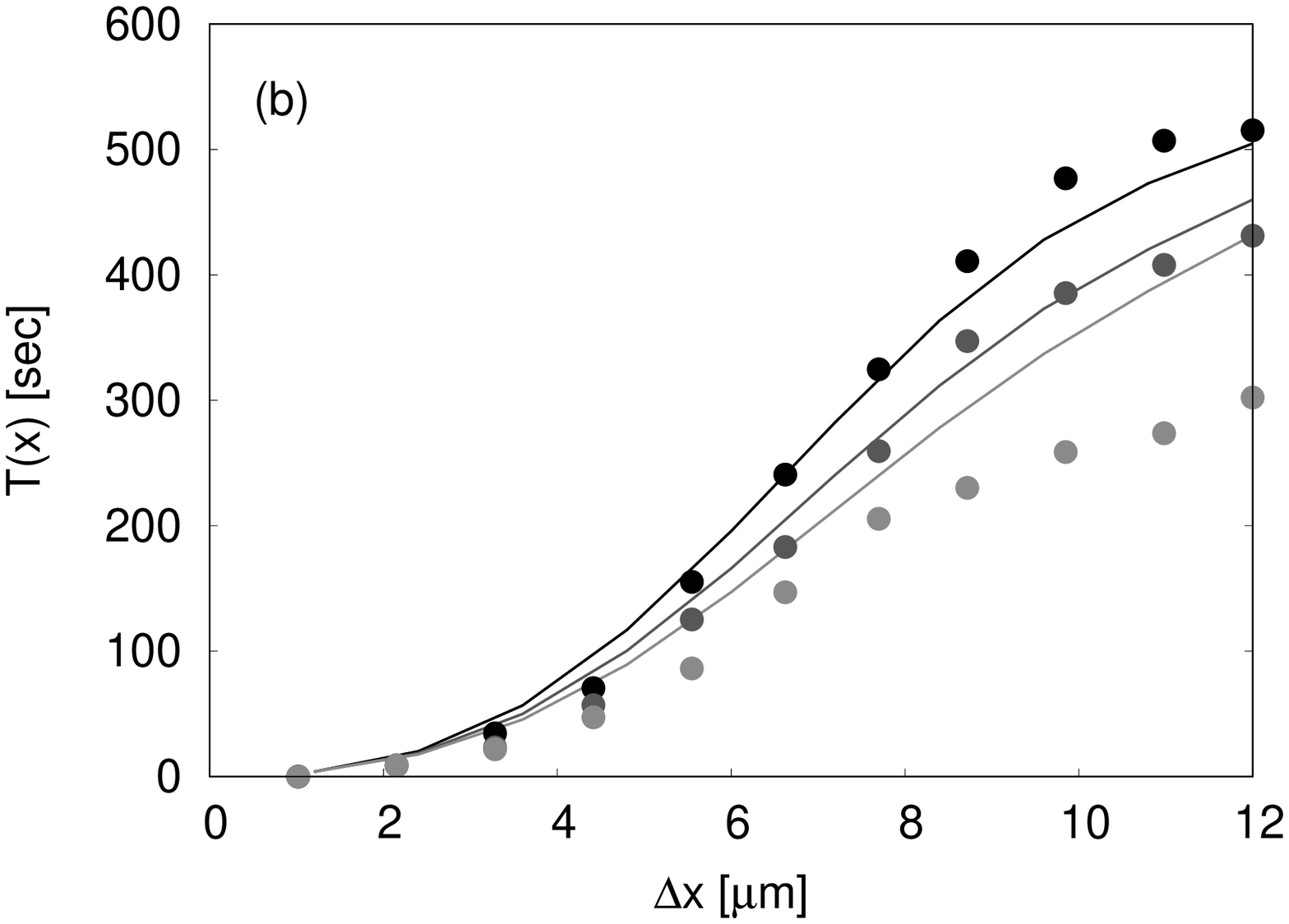}
\caption{Validation of the model (solid lines) against experimental data taken from Ref.~\cite{Hanggi2019} (dots).
(a) Dependence of the diffusion coefficient $\mathcal{D}(x)$ normalized
by the value at the bottleneck. In order to check the validity of
the model the values of the parameters are chosen to be close
to those used in Fig.~3a of Ref.~\cite{Hanggi2019}. Accordingly, we have:
$L_0=12\mu m$, $h_{0}=3\mu m$, $l=0.15\mu m$ and $L=1,1.2,1.6\mu m$. Lighter
colors stand for larger values of $L$. 
We remark the agreement between the model and the experimental data improves upon increasing the length of the rods. 
(b) Mean First Passage Time (MFPT) with reflecting boundary conditions at the origin and absorbing at $x$ for $L=1.04 \mu m,h_0= \mu m,h_1= \mu m$ (black),  $L=1.14 \mu m,h_0= \mu m,h_1= \mu m$ (dark green), $L=1.18 \mu m,h_0= \mu m,h_1= \mu m$ (light grey).
As shown in the panel, the agreement improves upon increasing the  corrugation of the channel.
}
\label{fig:Hanggi}
\end{figure*}
We remark that $p(x)$ (see Eq.~\eqref{eq:FJ0}) is proportional to the probability of finding the center of mass at a position $x$.
In the frame of reference of the rod we have 
\begin{equation}
\bar{\mathbf{D}}=D_{0}\left[\begin{array}{cc}
\frac{l}{L} & 0\\
0 & 1
\end{array}\right],
\end{equation}
where $D_0\frac{l}{L}$ is the diffusion coefficient along the minor axis of the rod (of size $l$), $\mathbf{e}_\perp$, and $D_0$ is the diffusion coefficient along the major axis of the rod (of size $L$), $\mathbf{e}_\parallel$, and the off-diagonal terms are zero due to the axial symmetry  of the rod~\cite{Brennerbook}.
Integrating Eq.~\eqref{eq:Jx} along $z$ and $\theta$ and using Eqs.~\eqref{eq:Jz}-\eqref{eq:FJ0} we obtain
\begin{align}
\int\limits_{0}^{\pi}\!\!\int\limits_{-\infty}^{\infty}\!\!J_x \frac{dzd\theta}{2h_0\pi}=&\, \partial_{x}\int\limits_{0}^{\pi}\!\!\int\limits_{-\infty}^{\infty}\!\!\mathbf{e}_{x}\cdot\mathbf{D}(\theta)\cdot\mathbf{e}_{x}\left[\partial_{x}\rho+\rho\beta\partial_{x}W\right]\frac{dzd\theta}{2h_0\pi}\nonumber\\
=&\partial_x\left[ \mathcal{D}(x)\left(\partial_{x}p+p\beta\partial_{x}A \right)  \right]\,.
%\partial_{x}\left\{ \frac{1}{2\pi h_{0}}\int\limits_{0}^{\pi}\int\limits_{-\infty}^{\infty}\mathbf{e}_{x}\cdot\mathbf{D}(\theta)\cdot\mathbf{e}_{x}\frac{e^{-\beta W(x,z,\theta)}}{e^{-\beta A(x)}}dzd\theta\times\left[\partial_{x}p+p\beta\partial_{x}A\right]\right\} 
\label{eq:int_z-theta}
\end{align}
Here, 
\begin{equation}
\mathcal{D}(x)=\int\limits_{0}^{\pi}\int\limits_{-\infty}^{\infty}\mathbf{e}_{x}\cdot\mathbf{D}(\theta)\cdot\mathbf{e}_{x}\frac{e^{-\beta W(x,z,\theta)}}{e^{-\beta A(x)}}\frac{dzd\theta}{2h_0\pi}\label{eq:def_Diff}
\end{equation}
is the effective \textit{local} diffusion coefficient.
 According to Fig.~\ref{fig:1} we have $\mathbf{e}_x=\cos\theta \mathbf{e}_\parallel - \sin\theta \mathbf{e}_\perp$ which leads to the following expression for the local diffusion coefficient,
\begin{equation}
\mathbf{e}_{x}\cdot\mathbf{D}(\theta)\cdot\mathbf{e}_{x}=D_{0}\left(\cos^{2}\theta+\frac{l}{L}\sin^{2}\theta\right)
\end{equation}
which, once substituted into Eq.~\eqref{eq:def_Diff}, leads to
\begin{equation}
\dfrac{\mathcal{D}(x)}{D_0}=\int\limits_{0}^{\pi}\int\limits_{-\infty}^{\infty}\left(\cos^{2}\theta+\frac{l}{L}\sin^{2}\theta\right)\frac{e^{-\beta W(x,z,\theta)}}{e^{-\beta A(x)}}\frac{dzd\theta}{2h_0\pi}\,.
\label{eq:def_Diff-1}
\end{equation}
We remark that for $l=L$ Eq.~\eqref{eq:def_Diff-1} reduces to $\mathcal{D}(x)=D_0$, which is the diffusion coefficient of spherical particles of size $l$~\cite{Zwanzig,Reguera2001}.
%whereas for $L\gg h_0$ we get $\mathcal{D}(x)\simeq D_0/2$.
Finally, integrating Eq.~\eqref{eq:smol-1} along $z$ and $\theta$ and using Eqs.~\eqref{eq:int_z-theta} leads to
\begin{equation}
\partial_{t}p(x,t)=-\partial_{x}\left\{ \mathcal{D}(x)\left[\partial_{x}p(x,t)+\beta p(x,t)\partial_{x}A(x)\right]\right\} \label{eq:smol-2}.
\end{equation}
This is the extension of the Fick-Jacobs approximation to the case of rigid rods embedded in varying-section channels. 
The steady-state solution of Eq.~\eqref{eq:smol-2} reads
\begin{equation}
p(x)=e^{-\beta A(x)}\left[J\int_{0}^{x}\frac{e^{\beta A(x')}}{\mathcal{D}(x')}dx'+\Pi\right]
\end{equation}
where $J$ and $\Pi$ are integration constants.
For periodic boundary conditions and imposing the normalization of
the probability we obtain
\begin{eqnarray}
\Pi & = & -J\left[\frac{\int_{0}^{L}\frac{e^{\beta A(x)}}{\mathcal{D}(x)}dx}{e^{-\beta(A(0)-A(L))}-1}\right]=-J \Pi_0\\
J & = & -\left[\int_{0}^{L}\!\!\!\!\!e^{-\beta A(x)}\left[\int_{0}^{x}\frac{e^{\beta A(x')}}{\mathcal{D}(x')}dx'+\Pi_{0}\right]dx\right]^{-1}\!\!\!\!\!\!.
\end{eqnarray}
%\jh{$\Pi$, ($J$)?} 
Finally we define the dimensionless permeability as
\begin{align}
    \mu=\frac{J}{J_0}=\frac{J}{\beta D_0 f/L},
    \label{eq:mu}
\end{align}
where $f$ is the magnitude of the external force and $J_0=\beta D_0 f/L$ is the flux of a spherical colloid of radius $l$ in a flat channel of half-width $h_0$.

% In particular the last expression can be rewritten as
% \begin{equation}
% \mathcal{D}(x)=\dfrac{\int\limits_{0}^{h(x)}f(\theta_{min}(x,z),\theta_{max}(x,z))dz}{\frac{2}{h_{0}}\int\limits_{0}^{h(x)}\left(1-\frac{2}{\pi}\arccos\left(\frac{h(x)-z}{l}\right)\right)dz}
% \end{equation}
% with
% \[
% f(\theta_{min}(x,z),\theta_{max}(x,z))=\frac{1}{L_{max}}\left(\cos\theta_{min}-\cos\theta_{max}\right)+\frac{1}{L_{min}}\begin{cases}
% 2-\left(\sin\theta_{min}+\sin\theta_{max}\right) & \theta_{min}<\frac{\pi}{2}\\
% \sin\theta_{min}-\sin\theta_{max} & \theta_{min}\geq\frac{\pi}{2}
% \end{cases}
% \]

\section{Results}

\subsection{Neutral rods}
At first we focus on neutral rods for which the  potential $W$ reads
\begin{equation}
W(x,z)=\begin{cases}
-fx & |z|<h(x)\,\&\,\theta_{m}<\theta<\theta_{M}\\
\infty & \mbox{else}
\end{cases}
\end{equation}
and the local free energy reads
\begin{align}
  A(x) & =-fx-k_{B}T\ln\left[\int\limits_{0}^{\pi}\!\int\limits_{-h(x)}^{h(x)}\!\!\Gamma(x,z,\theta)\frac{dzd\theta}{2h_0\pi}\right]\label{eq:Frod},
  \end{align}
with
\begin{align}
&\Gamma(x,z,\theta)=\Theta\left(\theta-\theta_{m}(x,z)\right)\Theta\left(\theta_{M}(x,z)-\theta\right),
\end{align}
where $\Theta(\theta)$ is the Heaviside step function.
%\begin{align}
%A(x) & =-fx-k_{B}T\ln\left[\frac{2}{h_{0}}\int\limits_{0}^{h(x)}\!\!\!\pi+\theta_{min}(z)-\theta_{max}(z)dz\right]\label{eq:Frod}
%\end{align}
Accordingly, the effective diffusion coefficient reads
\begin{align}
\dfrac{\mathcal{D}(x)}{D_0}=\frac{\int\limits_{0}^{\pi}\int\limits_{-h(x)}^{h(x)}\!\!\left[\cos^{2}\theta+\frac{l}{L}\sin^{2}\theta\right]\Gamma(x,z,\theta)dzd\theta}{\int\limits_{0}^{2\pi}\int\limits_{-h(x)}^{h(x)}\Gamma(x,z,\theta)dzd\theta},\label{eq:Diff}
\end{align}
% \begin{widetext}
%\begin{equation}
%\mathcal{D}(x)=\frac{D_{0}}{2\pi h_{0}}\dfrac{\int\limits_{0}^{2\pi}\int\limits_{-h(x)}^{h(x)}\left(\cos^{2}\theta+\frac{l}{L}\sin^{2}\theta\right)\Theta\left(\theta-\theta_{min}(x,z)\right)\Theta\left(\theta_{max}(x,z)-\theta\right)dzd\theta}{\int\limits_{0}^{2\pi}\int\limits_{-h(x)}^{h(x)}\Theta\left(\theta-\theta_{min}(x,z)\right)\Theta\left(\theta_{max}(x,z)-\theta\right)dzd\theta}\label{eq:Diff}
%\end{equation}
%\end{widetext}
where $L$ is the half-length of the major axis of the rod (see Fig.~\ref{fig:1}).
%\jh{Same L as before, isn't it? Check way you introduce it since I understood it to be the full-length of the major axis. L is not shown in Fig 1, btw., but it would be good to add it.}

\subsubsection{Diffusion}
\begin{figure}
\centering
\includegraphics[scale=0.35]{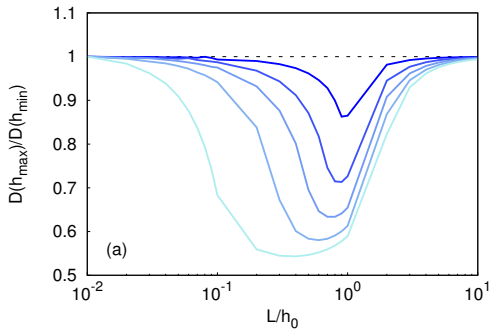}
\caption{Ratio of the effective diffusion coefficient at the maximum channel
amplitude ($D(h_{max})$) and at the bottleneck ($D(h_{min})$) as a function of $L$ and for different values of $\beta\Delta A_{gas}=0,0.2,0.6,1.1,1.7,2.9$ 
(from blue to cyan). Rods have a fixed magnitude of the small axis $l/h_0=0.01$.
} 
\label{fig:diff_z0}
\end{figure}
\begin{figure}
\includegraphics[scale=0.35]{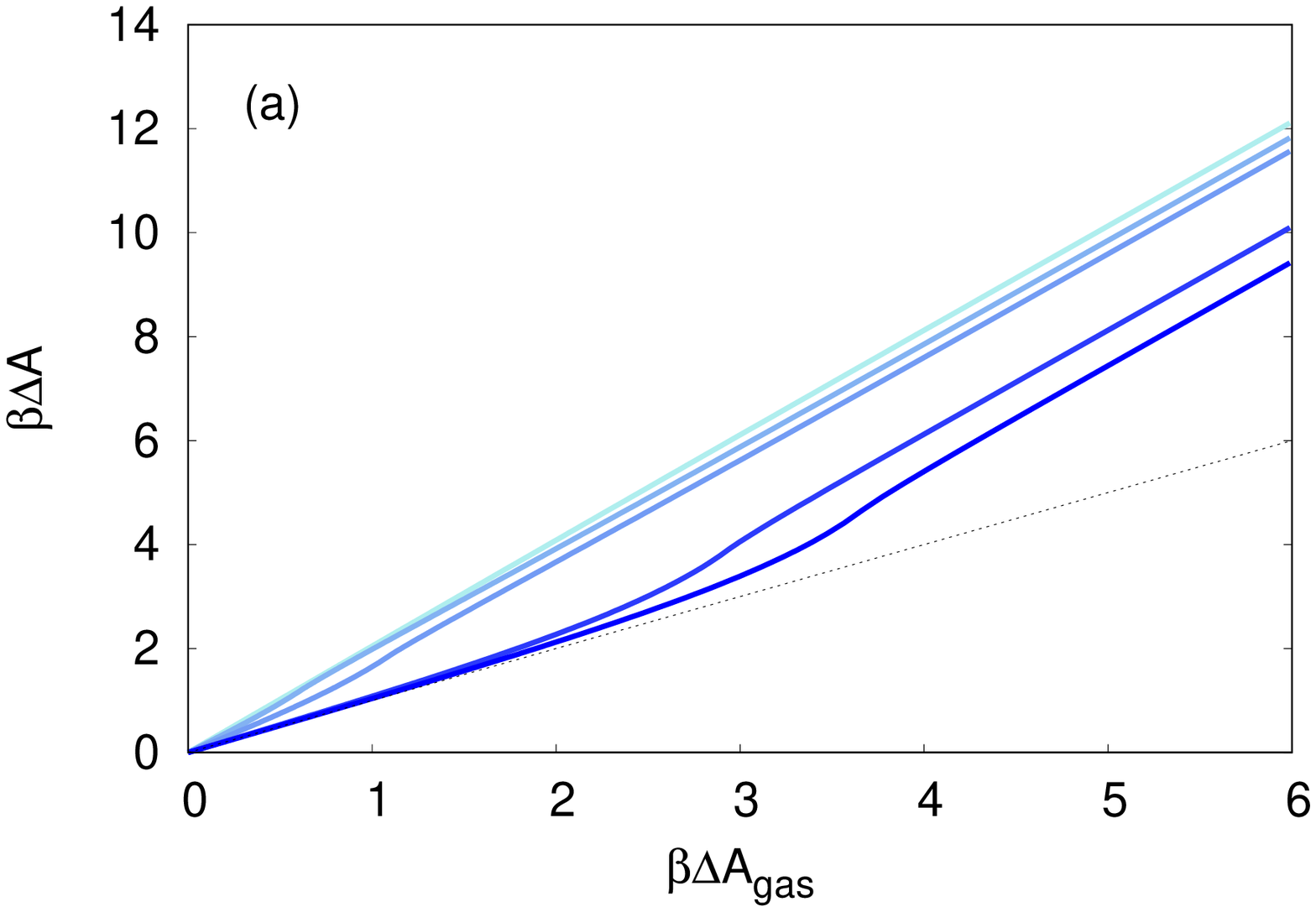}
\includegraphics[scale=0.35]{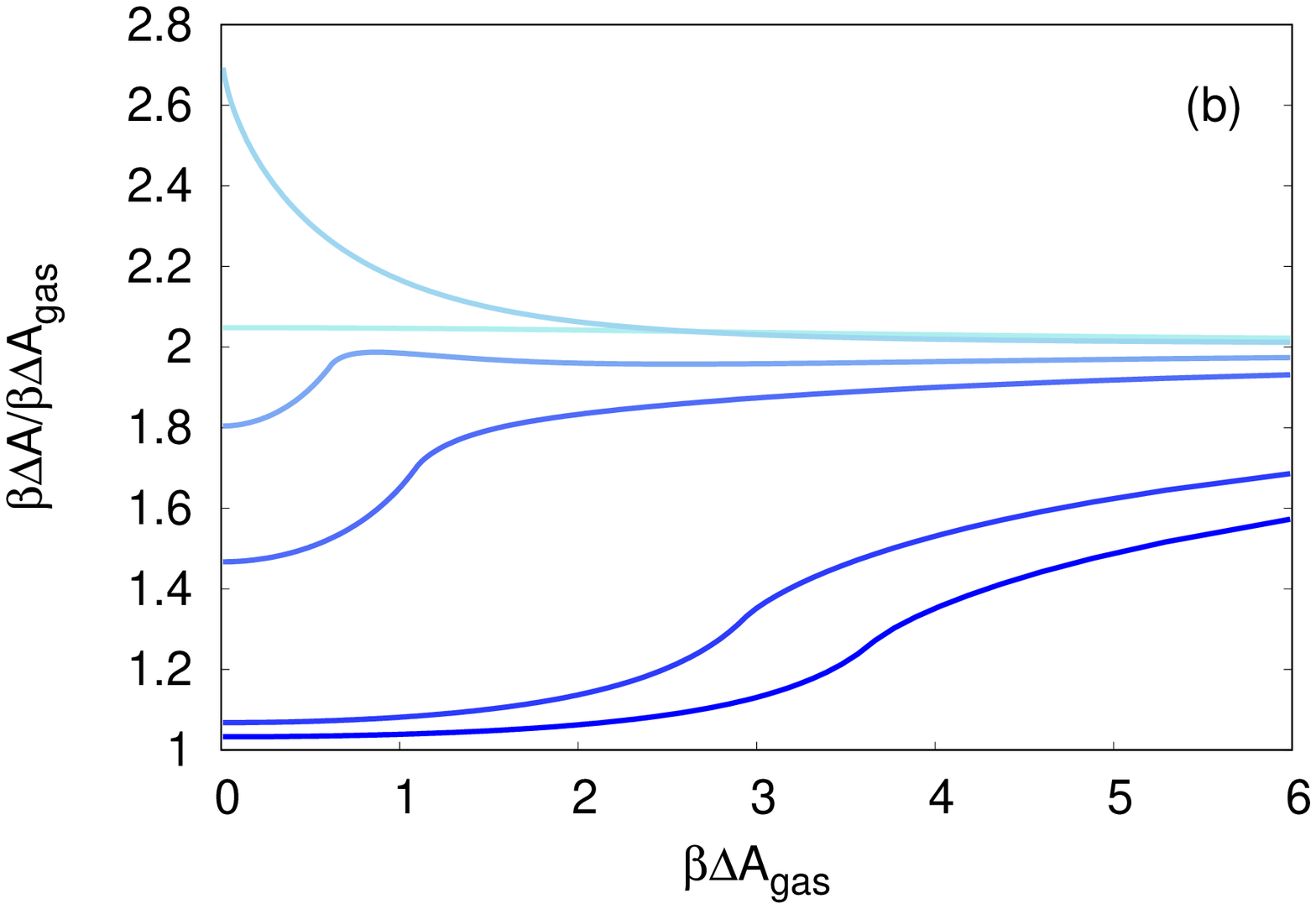}
\includegraphics[scale=0.35]{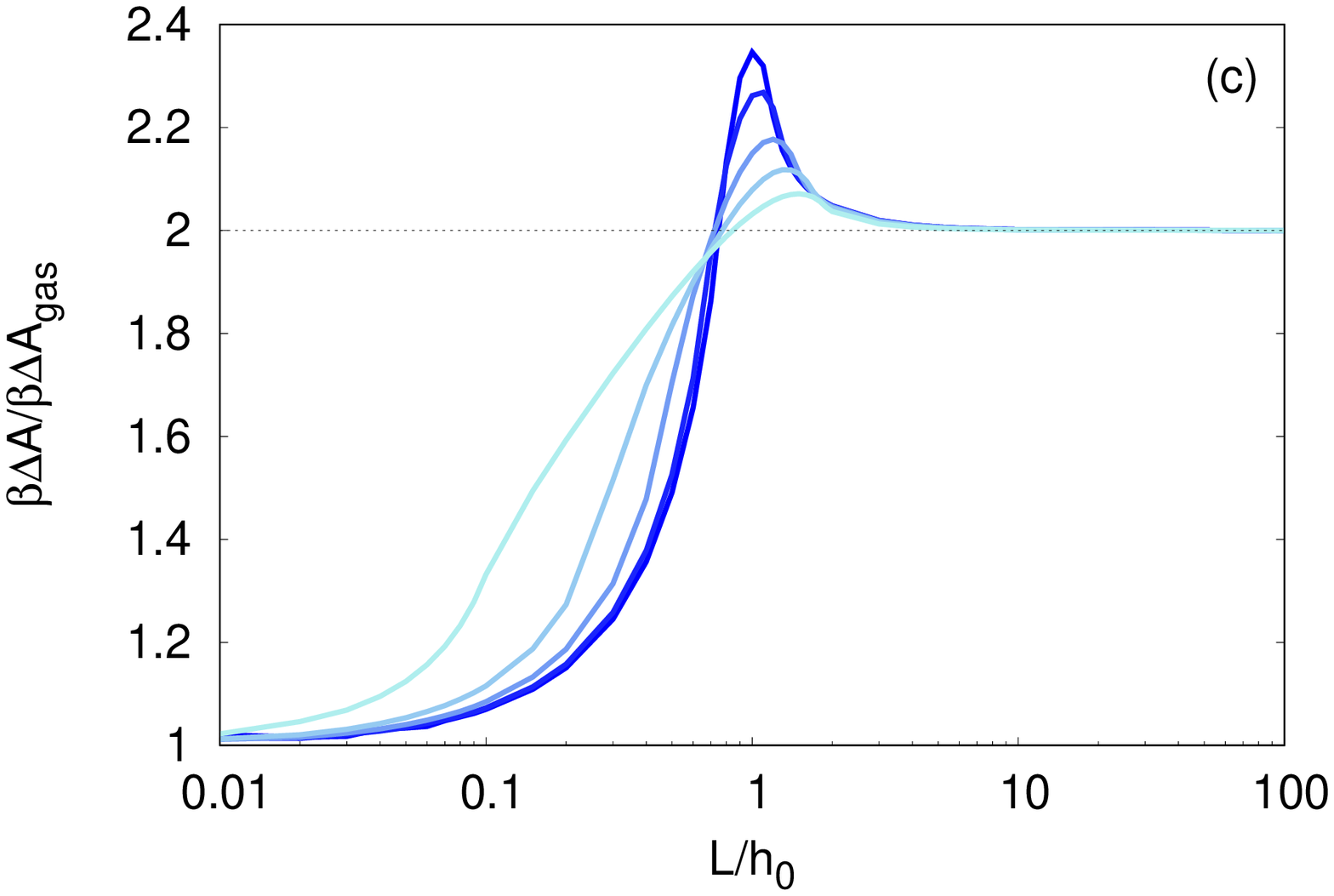}
\caption{Equilibrium free energy barrier $\Delta A$. (a) and (b) depict the dependence of the free energy barrier, $\Delta A$, for rods of different length
$L/h0=0.05,0.1,0.5,0.7,1,2$ (from blue to cyan) as a function of the
free energy barrier of the ideal gas. (c)  Dependence of the free energy barrier on the length, $L$, of the
rod for different channel corrugations $\Delta A_{gas}\simeq0.4,0.6,1.1,1.7,3$.}
\label{fig:Eq-DS}
\end{figure}
At first we compare our analytical predictions for the diffusion
coefficient against experimental results. 
As already shown in Ref.~\cite{Hanggi2019}, the diffusion coefficient
is maximum at the channel bottleneck where the rod is mainly parallel
to the axis of the channel. Fig.~\ref{fig:Hanggi}a
shows that predictions of Eq.~\eqref{eq:Diff} qualitatively agree with the experimental data. In particular, the model properly captures the enhancement of the sensitivity of the local diffusion coefficient upon increasing the length of the rod. Indeed, we remark that for longer rods, for which the dependence of the diffusion coefficient is more significant, the predictions of the model are quantitatively reliable, whereas the agreement with the experimental data becomes weaker when the dependence of $D$ on $x$ becomes milder.
%and the data from Ref.~\cite{Hanggi2019} is quite good \pa{ in the case of longer rods whereas it becomes worst in the case of shorter rods. We remark that in the case of shorter rods the variation of the diffusion coefficient is very mild However, even in the worst case the prediction is at most $10\%$ off.}
Once the agreement for the diffusion coefficient has been assessed we focus on the validity
of the Fick-Jacobs approximation for what concerns the MFPT between the origin and an arbitrary position $x$: 
\begin{equation}
T(x)=\int\limits_{0}^{x}dxe^{\beta A(x')}\int\limits_{0}^{x'}\frac{e^{-\beta A(z)}}{\mathcal{D}(z)}dz\label{eq:MFPT}
\end{equation}
Interestingly, the predictions of Eq.~\eqref{eq:MFPT} match well with
the data reported in Ref.~\cite{Hanggi2019} in the case of more corrugated channels whereas it is just qualitative for more shallow shapes. Interestingly, this trend is similar to that of the model (based on finite-element numerical simulations) reported in Ref.~\cite{Hanggi2019}.

Next, we characterize the dependence of
the ratio between the maximum and the minimum local diffusion coefficients
on the length of the rod. As expected, Fig.~\ref{fig:diff_z0} shows
that in the asymptotic limits, $L/h_{0}\ll1$ and $L/h_{0}\gg1$, the rods either reduce to a spherical particle of radius $l=L$ or to a very long stiff filament that, due to the confinement is almost always aligned with the channel axis. In both cases the effective diffusion coefficient reduces to $\mathcal{D}\simeq D_0$. In contrast, for $L\simeq h_{0}$,
Fig.~\ref{fig:diff_z0} %\jh{There is no 3a/3b, just 3.} 
shows that increasing the corrugation of the channel increases the range of values of the rod length for which
the local diffusion coefficient is sensitive to the local channel
section. Clearly, if the aspect ratio $L/l$ approaches unity the
diffusion coefficient becomes homogeneous along the channel. %Fig.~\ref{fig:diff_z0}b shows that the local diffusion coefficient, as a function of the aspect ratio, attains an asymptotic value already for $\frac{L}{l}\sim10$. 

\subsubsection{Free energy barrier}

Having an explicit formula for the free energy, it allows us to discuss
the dependence of the equilibrium ($f=0$) free energy barrier defined
as the difference between the free energy at the bottleneck and the
one at the channel's widest section:
\begin{equation}
\Delta A_{eq}=A(h_{min},f=0)-A(h_{max},f=0)
\label{eq:DA-eq}
\end{equation}
For comparison we recall that the free energy difference for an ideal
gas depends solely on the geometry of the channel:
\begin{equation}
\Delta A_{gas}=k_BT\ln\left[\frac{h_{max}}{h_{min}}\right]
\end{equation}
Our model shows that the dependence of the free energy barrier of rods strongly depends on their length. Indeed, while for shorter
rods, $L/h_{0}\lesssim0.5$, the free energy barrier of the rod increases
upon increasing channel corrugation (i.e. increasing $\Delta A_{gas}$)
for longer rods, $L/h_{0}\gtrsim1$, %\jh{L instead of l?}, 
the free energy barrier of the
rod decreases upon increasing $\Delta A_{gas}$, as shown in Fig.~\ref{fig:Eq-DS}.
Additionally, for every geometry of the channel  the maximum departure of $\Delta A$ from $\Delta A_{gas}$ is attained when the length of the rod is comparable to the channel average section,
$L\simeq h_{0}$. Interestingly, for rods much longer than the channel
section we have that $\Delta A\rightarrow2\Delta A_{gas}$. In particular, very long rods $L\gg h_0$ imply that $z\simeq 0$, i.e. that the center of mass is confined close to the channel axis. Hence, this asymptotic behavior can be understood by expanding Eq.~\eqref{eq:Frod} about $z=0$. Indeed, for $L/h_0\gg 1$ the integrand of Eq.~\eqref{eq:Frod} can be approximated by $h(x)/L$ and hence we obtain that $A(x)\propto2\ln\left[h(x)\right]$
which eventually leads to $\Delta A\rightarrow2\Delta A_{gas}$.

\subsubsection{Transport}

\begin{figure}
\centering
\includegraphics[scale=0.4]{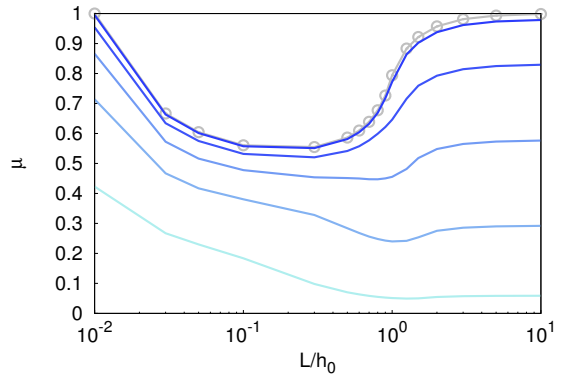} \caption{Dimensionless channel permeability, $\mu$, to non-charged rods as function of the normalized length of the major axis of the rods, $L/h_0$ for rods with  $l=10^{-2}h_0$ and for different values of  $\beta\Delta A_{gas}=0.2,0.6,1.1,1.7,2.9$ (from blue to cyan), whereas the grey dots are for $\beta\Delta A_{gas}=0$.
\label{fig:flux_z0}}
\end{figure}

Next, we analyze the transport of rods under the action of a constant
force. In order to simplify the analysis we assume that the fluid rods are suspended and keep at rest, i.e. there is no advection of rods due to fluid motion. 
%To this regard is insightful to look at the channel permeability:
%\begin{align}
% \mu=\frac{J L_0}{\beta \bar{\mathcal{D}}(L) f}\,,
%\end{align}
%where 
%\begin{align}
% \hat{\mathcal{D}}(L)=\frac{1}{L}\int\limits_{0}^{L}\mathcal{D}(x,L)dx
% \label{eq:D-hat}
%\end{align}
%is the diffusion coefficient averaged over the channel period, $L$.
In such a regime, Fig.~\ref{fig:flux_z0} shows that the dependence
of the dimensionless permeability $\mu$ on the length of the major axis of the rod, $L$, is quite complex. Indeed, for flat channels ($\beta \Delta A_{gas}=0$) the transport is determined solely by the effective diffusion coefficient. In particular,  for $L\gg h_0$ the rod is almost aligned with the axis of the channel and the effective diffusion coefficient approaches the one of the minor axis (see Fig.\ref{fig:diff_z0}) and, according to Eq.\eqref{eq:mu} $\mu\rightarrow 1$ for $L\gg h_0$. In contrast, for $\beta \Delta A_{gas}=0$ and $L\ll h_0$ the rod can freely rotate and it experiences a reduction in the effective diffusion coefficient (see Fig.\ref{fig:diff_z0}).
For $\beta \Delta A_{gas}\neq 0$, the channel is not flat and the dependence of the dimensionless permeability $\mu$ on $L$ becomes more involved. Indeed, Fig.~\ref{fig:flux_z0} shows that for intermediate values of $\beta \Delta A_{gas}$, $\mu$ displays a non-monotonous dependence on $L/h_0$. This behavior is similar to that observed for polymers confined between corrugated plates (see Ref.~\cite{Bianco2016}). Finally, for larger values of $\beta \Delta A_{gas}$, $\mu$ monotonically decreases upon increasing $L/h_0$.

% \begin{figure}
% \includegraphics[scale=0.4]{flux_z}
% \caption{Flux of a charged rod, $J$, across a corrugated channel, normalized
% by the flux of the same rod in a flat channel, $J_{0}$, for different
% values of the channel corrugation, $h_{1}/h_{0}=0.5,0.7,0.8,0.9$,
% as a function of $\beta q \phi_0$, with $\kappa h_{0}=1$. Lighter colors stand
% for larger values of $h_{1}.$ The force is independent of the charge
% and amounts to $\beta fL=0.1$.}
% \label{fig:flux_z}
% \end{figure}
\subsection{Charged rods}

Next, we analyze the case of charged rods. For simplicity we assume
that the charge, $q$, is localized at the center of the rod and
that the channel walls are characterized by a constant charge density 
$\sigma$. In addition, we assume that a dilute monovalent electrolyte
is suspended in the fluid phase so that the system is electrically
neutral. In this regard, the rod brings an extra charge, whose magnitude
is assumed to be much smaller than that of the charge of the double layer such that it does not affect significantly the local electrostatic field\footnote{Such an approximation becomes exact in the limit of vanishing charge of the rod.}.
%\jh{Is this a realistic/reasonable assumption?}
%\pa{The idea is that the double layer is composed of many elementary charges whereas the rod will carry only a few (or even just one).}
%\jh{I understood that. But is this something one can expect to be the case in reality? I guess not...}
%\pa{I think that we should not enter in these details here. In principle, one should solve the Debye-Huckel equation in the presence of the rod, but this is not possible analytically. However, in the limit of small charge of the rod, I think that it is reasonable to assume that the perturbation of the electric field will be small. Actually this is exact for zero charge of the rod. I have added a footnote in which I mention this.} %In particular such an estimation should be better for rods shorter than the Debye length for which the cloud of counterions about the rod is very "extended" as compared to its own size.}
Accordingly, the (unperturbed) electrostatic potential inside the channel reads
\begin{equation}
\phi=\frac{\sigma}{\varepsilon\kappa}\frac{\cosh\left(\kappa z\right)}{\sinh\left(\kappa h\left(x\right)\right)}=\phi_0\frac{\cosh\left(\kappa z\right)}{\sinh\left(\kappa h\left(x\right)\right)}\,,
\end{equation}
with $\phi_0=\frac{\sigma}{\varepsilon\kappa}$. %In the following we fix $\phi_0=1$. 
The local free energy becomes
\begin{align}
  A(x) & =-fqx-k_{B}T\ln\left[\int\limits_{0}^{\pi}\!\int\limits_{-h(x)}^{h(x)}\!\!e^{-\beta q\phi(x,z)}\Gamma(x,z,\theta)\frac{dzd\theta}{2h_0\pi}\right]\label{eq:Frod-charge},  
\end{align}
where %\jh{repeated:} \sout{$q$ is the charge of the rod and} 
the diffusion coefficient reads
\begin{equation}
\dfrac{\mathcal{D}(x)}{D_0}=\dfrac{\int\limits_{0}^{\pi}\int\limits_{-h(x)}^{h(x)}\!\!\left[\cos^{2}\theta+\frac{l}{L}\sin^{2}\theta\right]e^{-\beta q\phi(x,z)}\Gamma(x,z,\theta)dzd\theta}{\int\limits_{0}^{\pi}\int\limits_{-h(x)}^{h(x)}e^{-\beta q\phi(x,z)}\Gamma(x,z,\theta)dzd\theta}.
\label{eq:Diff_charge}
\end{equation}
%\begin{equation}
%A(x)=-fzex-k_{B}T\ln\left[\frac{2}{h_{0}}\int\limits_{0}^{h(x)}e^{-\beta q\phi_{0}\frac{\cosh\left(\kappa z\right)}{\sinh\left(\kappa h\left(x\right)\right)}}\left(\pi-2\arccos\left(\frac{h(x)-z}{L}\right)\right)dz\right]\label{eq:Frod-charge}
%\end{equation}
%\begin{equation}
%\mathcal{D}(x)=\frac{D_{0}}{2\pi h_{0}}\dfrac{\int\limits_{0}^{2\pi}\int\limits_{-h(x)}^{h(x)}\left(\cos^{2}\theta+\frac{l}{L}\sin^{2}\theta\right)e^{-\beta q\phi_{0}\frac{\cosh\left(\kappa z\right)}{\sinh\left(\kappa h\left(x\right)\right)}}\Theta\left(\theta-\theta_{min}(x,z)\right)\Theta\left(\theta_{max}(x,z)-\theta\right)dzd\theta}{\int\limits_{0}^{2\pi}\int\limits_{-h(x)}^{h(x)}\Theta\left(\theta-\theta_{min}(x,z)\right)\Theta\left(\theta_{max}(x,z)-\theta\right)dzd\theta}\label{eq:Diff_charge}
%\end{equation}
\begin{figure}
 \centering
\includegraphics[scale=0.4]{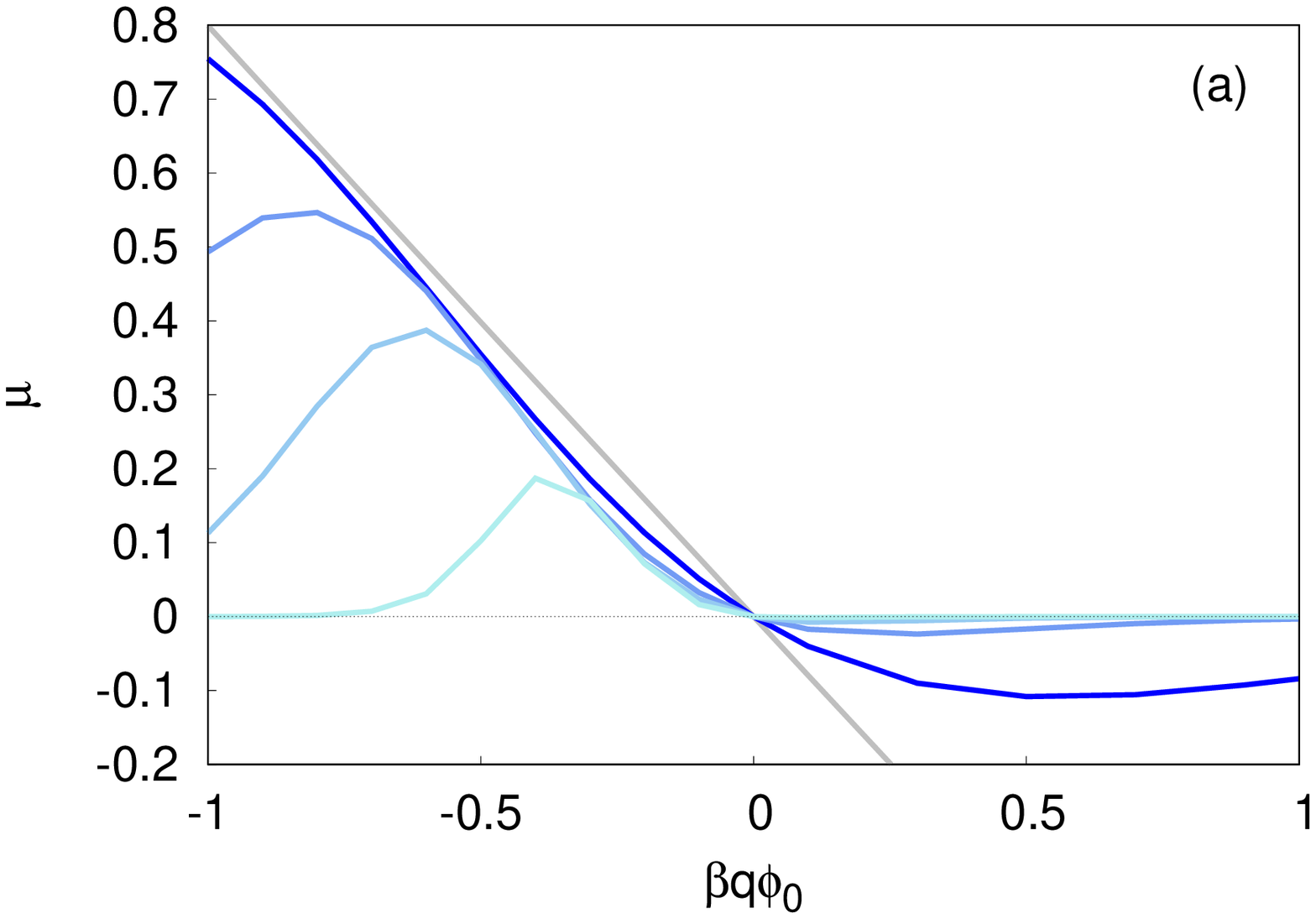}
\includegraphics[scale=0.4]{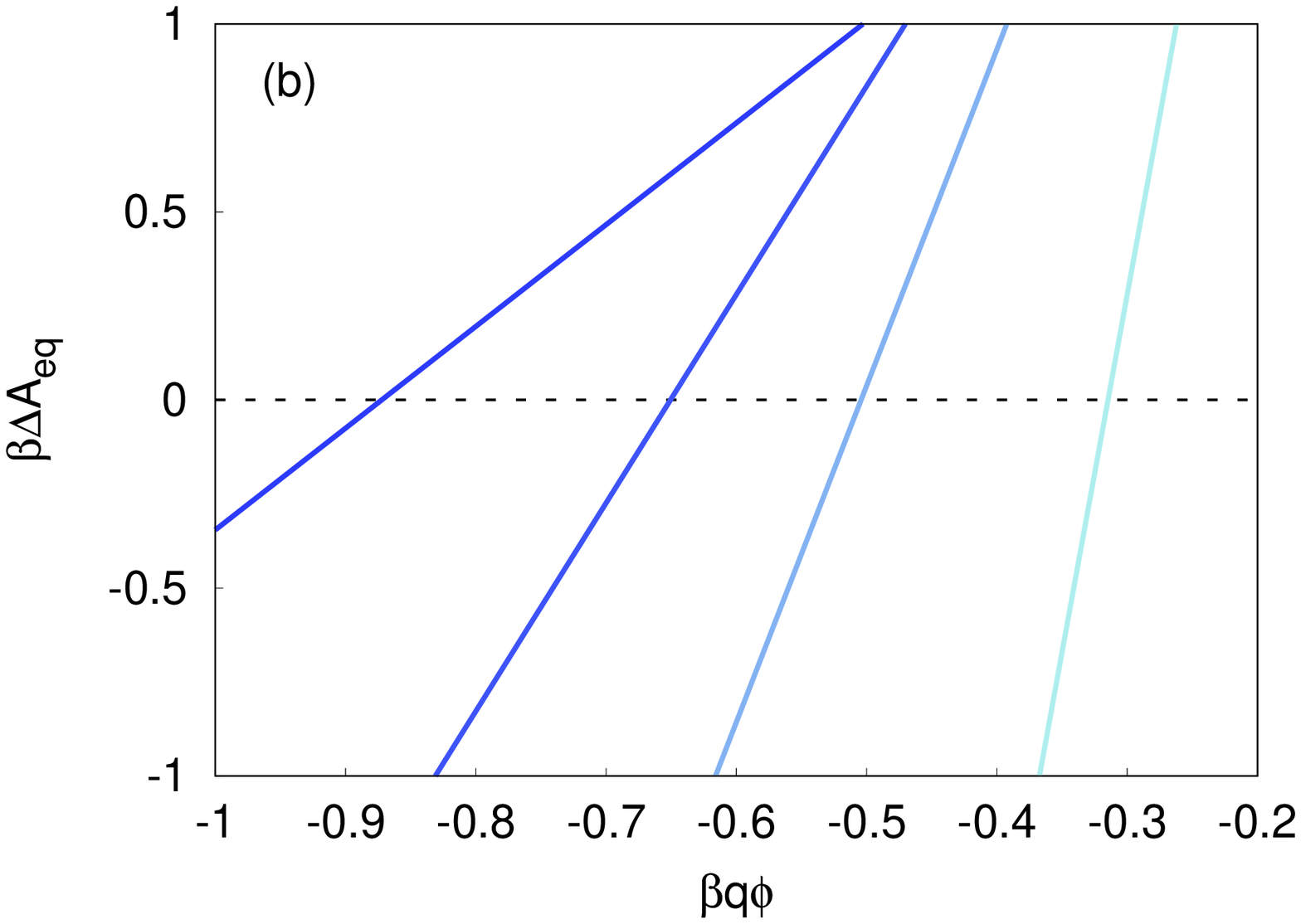}
\caption{Dimensionless channel permeability, $\mu$, as function of the charge of the rods.
(a): $\mu$ as a function of $\beta q \phi_0$ for different
values of the channel corrugation, $\beta\Delta A_{gas}=1.1,1.7,2.2,2.9$, with $\kappa h_{0}=1$. Lighter colors stand
for larger values of $\beta\Delta A_{gas}$, whereas the black solid line stands for $\beta\Delta A_{gas}=0$. The force is proportional to the charge and, for a unit charge, it amounts to $\beta f e L=0.1$.
(b): $\beta\Delta A_{eq}$ (see Eq.~\eqref{eq:DA-eq}) as function of the interaction potential with the walls $\beta q\phi_0$. Note that the maxima in panel (a) occur for values of $q$ for which $\beta\Delta A_{eq}\simeq0$.
}
\label{fig:flux_charge-1}
\end{figure}
%\noindent and that the magnitude of the external force is not depending on the value of the charge, $ze$. 
The dependence of the channel permeability on the electrostatic interaction of the rods with the walls is shown in Fig.~\ref{fig:flux_charge-1}a. In particular, Fig.~\ref{fig:flux_charge-1}a shows that the dependence of $\mu$ on the interaction potential of the rods with the walls is non-monotonous. For a given charge density on the walls, when the magnitude of the charge is very large (for both signs of the charge) the permeability becomes vanishingly small. Indeed, for such cases the effective free energy barrier to be overcome is very large and hence $\mu$ becomes very small. In contrast, upon reducing the magnitude of the charge, $\mu$ increases and it reaches its maximum for weakly negatively charged rods, i.e. when rods are \textit{attracted} by the channel walls. As shown in Fig.~\ref{fig:flux_charge-1}a the optimal value of the charge depends on the channel corrugation and it decreases upon increasing the channel corrugation. 
This non-monotonous dependence, and in particular the location of the maximum, can be understood by looking at the equilibrium free energy difference $\beta\Delta A_{eq}$. Indeed, Fig.~\ref{fig:flux_charge-1}b  shows that the maxima observed in Fig.~\ref{fig:flux_charge-1}a occur for values of $q$ for which $\beta \Delta A_{eq}=0$ (see Fig.~\ref{fig:flux_charge-1}b) i.e., when the equilibrium overall free energy barrier drops and the rods have only to overcome smaller local free energy barriers.
%\subsubsection{Particle separation}

Fig.~\ref{fig:flux_charge-1}a shows that the geometry of the channel can be used to tune the sorting of rods depending on their charge. In order to exploit the geometry of the channel and to separate charged rods depending on their length, 
it is mandatory to find the set of parameters for which the sensitivity of the dimensionless channel permeability on the channel geometry is maximized. As discussed, our model highlights a direct correlation between the equilibrium free energy barrier $\Delta A_{eq}$ and the dimensionless channel permeability $\mu$. This implies that the region of the parameter space in which $\mu$ is likely to be sensitive to the geometry, $\Delta A_{gas}$, can be guessed by inspecting the equilibrium free energy difference $\Delta A_{eq}$. 
\begin{figure}[t!]
    \centering
    \includegraphics[scale=0.38]{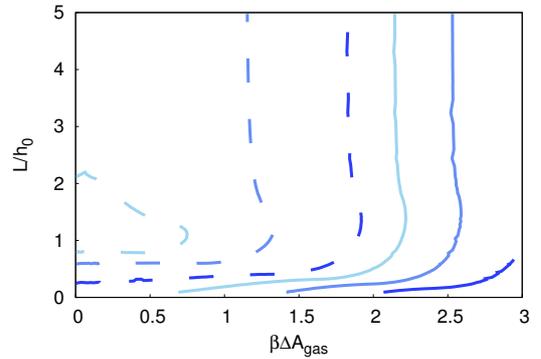}
    \caption{Contour lines of $\beta \Delta A_{eq}=0$ for  $kh_0=0.5$ (dashed lines) and $kh_0=1$ (solid lines). Different colors are for diverse values of $\beta q \phi_0=-0.6,-0.8,-1$  where lighter colors   stand for larger values of the absolute value of  $\beta q \phi_0$.}
    \label{fig:contour_DF}
\end{figure}
\begin{figure}[t!]
    \centering
    \includegraphics[scale=0.38]{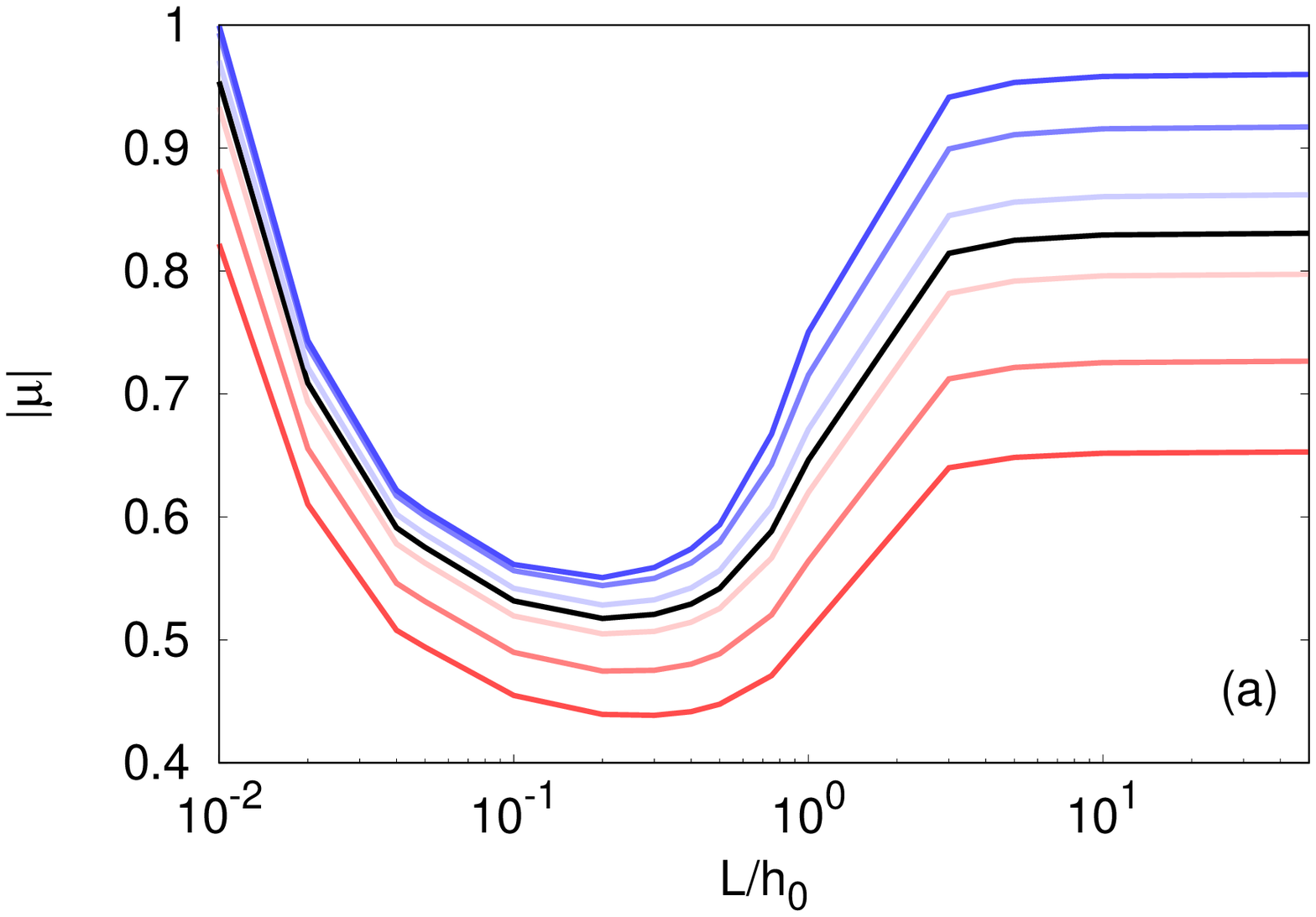}
    \includegraphics[scale=0.38]{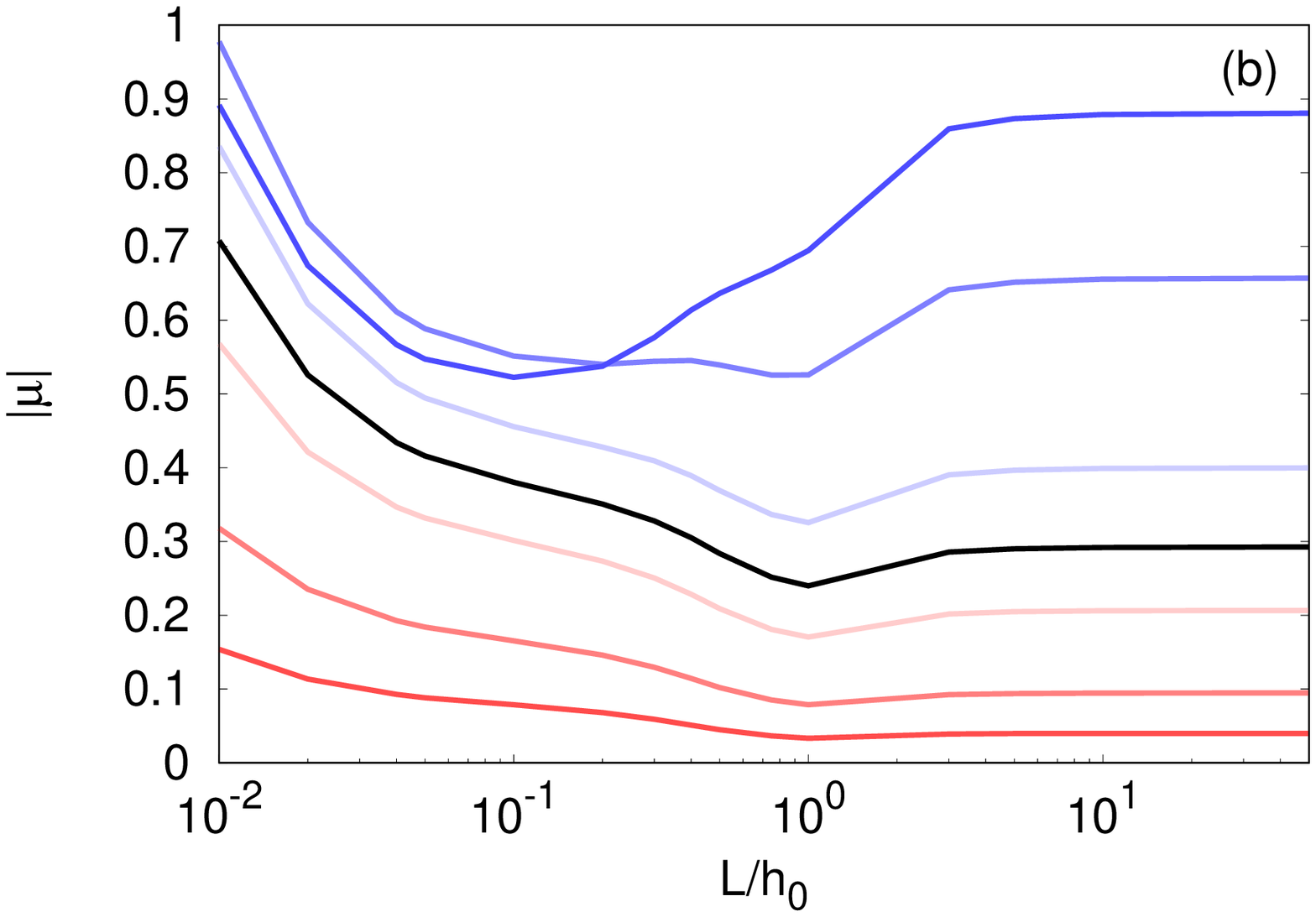}
    \caption{Absolute value of the dimensionless channel permeability, $\mu$, to  charged rods as function of the normalized length of the major axis of the rods, $L/h_0$, for $\kappa h_{0}=0.5$, $\beta\Delta A_{gas}=1.7$ (panel (a)), $\beta\Delta A_{gas}=0.6$ (panel (b)), $L/l=100$ and for different values of the wall potential, $\beta q\phi_{0}=-0.5,-0.3,-0.1,0,0.1,0.3,0.5$, that
is color coded: blue (red) lines stand for negative (positive) charges
and lighter colors stands for smaller magnitudes of the wall potential. The
black solid line stands for $z=0$. All results are for $\beta q f L=0.1$. }
\label{fig:mu_star}
\end{figure}
\begin{figure}[t!]
\centering
\includegraphics[scale=0.38]{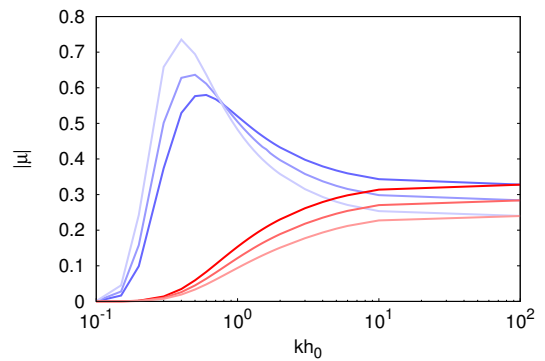}
\caption{Absolute value of the dimensionless channel permeability, $\mu$, to charged rods upon varying the inverse dimensionless Debye length, 
%(a): $\mu$ as a function of the channel corrugation encoded in $\Delta A_{gas}$ for $\kappa h_{0}=1$ , $L/h_{0}=1$, $L/l=100$ and for different values of the wall potential, $\beta q\phi_{0}=-0.5,-0.3,-0.2,-0.1$, in blue colors lighter colors stands for smaller magnitudes of the charge, $\beta q\phi_{0}=0$, in black and $\beta q\phi_{0}=0,0.1,0.3,0.5$ in red colors lighter colors stands for smaller magnitudes of the wall potential. 
%(b): $\mu$ as a function of the length of the major axis of the rod, $L$, normalized by $h_0$ for $\kappa h_{0}=1$, $h_{1}=0.7$, $L/l=100$ and for different values of the wall potential, $\beta q\phi_{0}=-0.5,-0.3,-0.1,0,0.1,0.3,0.5$, that is color coded: blue (red) lines stand for negative (positive) charges and lighter colors stands for smaller magnitudes of the wall potential. The black solid line stands for $z=0$. 
$\kappa h_0$, for $\beta\Delta A_{gas}=1.7$,
$L/l=100$ and for  $\beta q\phi_{0}=-0.5$ (blue lines) and  $\beta q\phi_{0}=0.5$ (red lines). 
In blue colors lighter colors stands for larger values of $L/h_{0}=0.3,0.5,1$
and $\beta q\phi_{0}=0.5$ in red colors lighter colors stands for
larger values of $L/h_{0}=0.3,0.5,1$. All results are for $\beta q f L=0.1$.}
\label{fig:flux_charge-2}
\vspace{-5pt}
\end{figure}
Fig.~\ref{fig:contour_DF} shows that for $kh_0\gtrsim1$, the isolines $\beta\Delta A_{eq}=0$ are parallel to the ordinate axis for $L\gtrsim h_0$. Hence, all rods with length $L\gtrsim h_0$ are expected to have relatively similar values of $\mu$. Accordingly, for this set of parameters only rods with length $L\lesssim h_0$ can be separated. In contrast, for $kh_0\simeq 0.5$ and for $\beta \Delta A_{gas}\simeq 0.75$ the contour line $\beta \Delta A_{eq}=0$ bends. This can be the signal of a strong dependence of the velocity on the rod size. 
%Since the diffusion coefficient depends on the rod longer axis $L$, the dimensionless permeability $\mu$ is not a good observable when discussing rod  separation according to their length since, in $\mu$ both the numerator and the denominator depend on $L$. Accordingly, we define 
%\begin{align}
%    \mu^*=\dfrac{J}{D_0 f/L}=\mu \dfrac{\hat{\mathcal{D}}}{D_0}
%    \label{eq:mu-star}
%\end{align}
%where $D_0$ is the diffusion coefficient of a spherical particle of size $l$ and $\hat{\mathcal{D}}$ is defined in Eq.~\eqref{eq:D-hat}.
Fig.~\ref{fig:mu_star}a shows that for larger values of $\beta \Delta A_{gas}$ of and $\beta q \phi_0\gtrsim -0.1$, $\mu$ is sensitive to $L/h_0$ only for $L\lesssim h_0$ whereas $\mu$ shows a plateau for $L\gtrsim h_0$, in agreement with our argument based on $\beta \Delta A_{eq}$.
In contrast, for smaller values of $\beta \Delta A_{gas}$, Fig.~\ref{fig:mu_star}b shows that  for $\beta q \phi_0\lesssim -0.1$, $\mu$ is more sensitive to $L$ for the full range of values explored in Fig.~\ref{fig:mu_star}b, again in agreement with our argument.

%dependence of  $\mu$ on the channel corrugation, encoded in $\Delta A_{gas}$, for diverse values of the charge of the rod. Interestingly, we notice that for positively charge rods, i.e. repelled from the channel walls,  $\mu$ is hampered as compared to the neutral charge case. In contrast, negatively charged rods, hence attracted by the channel walls, experience an enhanced permeability as compared to the neutral charged ones. 
%This scenario is persists for all rod lengths we have investigated. Fig.~\ref{fig:flux_charge}.c shows that the dependence of  $\mu$ on the length of the rod is similar to that of neutrally charged rods, given the enhancement (reduction) factor for negatively (positively) charge rods that we have discussed above. 
Finally we have studied the dependence of the flux of charged rods as a function of the Debye length. Interestingly, Fig.~\ref{fig:flux_charge-2}  shows that the flux of negatively charged rods (hence attracted by the channel walls) has a non-monotonous dependence on the dimensionless inverse Debye length $\kappa h_0$ and it displays a maximum around $\kappa h_0\lesssim 1$ for all the geometries of the channel that we have explored. In contrast, the net flow of positively charged rods increases monotonously with the inverse Debye length $\kappa h_0$.

\subsection{Conclusions}
We have studied the dynamics of charged rods embedded in varying-section channels. 
Under the assumption of slowly varying channel sections, $\partial_x h(x)\ll 1$ we have extended the Fick-Jacobs approximation to the case of charged rods. Our approximation allows us to derive an expression for the local diffusion coefficient, $\mathcal{D}(x)$ (see Eq.~\eqref{eq:def_Diff-1}). 
We have tested our prediction for the case of neutrally charged rods against experimental and numerical results (see Ref.~\cite{Hanggi2019}). Interestingly, our predictions (see Fig.~\ref{fig:Hanggi}) match well with the experimental/numerical data for both the local diffusion coefficient as well as for the Mean First Passage Time. We remark that, while in experiments hydrodynamic coupling between the rods and the channel walls is naturally accounted for, it is not so in our model.  Hence, a first result of our analysis is that, in the regime under study, hydrodynamic coupling plays a minor role in the diffusion of rods within corrugated channels.

In order to grasp the relevance of the rod geometry on the free energy barrier, $\Delta A$, we have compared it  to that of point particles, $\Delta A_{gas}$. Interestingly, we found that the enhancement in the free energy difference, $\Delta A$, is at most twofold (see Fig.~\ref{fig:Eq-DS}c).
Next, we have characterized the channel permeability, $\mu$, to neutral rods as a function of their length and 
found that for intermediate values of $\beta \Delta A_{eq}$, $\mu$ displays a non-monotonous dependence (see Fig.~\ref{fig:flux_z0}) on the major axis of the rod, $L$, similarly to what has been observed for confined polymers~\cite{Bianco2016}.

Finally, we have characterized the channel permeability to charged rods. In particular, we have focused on the dependence of $\mu$ on the rod size and charge. Interestingly, we found that $\mu$ displays a non-monotonous dependence on the rod charge and that the value of the charge that maximizes $\mu$ is the one for which the overall equilibrium free energy barrier, $\Delta A_{eq}$ vanishes. 
Hence, our model allows us to rationalize the non-monotonous dependence of $\mu$ with respect to the charge of the rod.
We have exploited this feature to 
%Hence, by inspecting the isolines $\beta \Delta A=0$ and we have been able to 
identify those set of parameters for which $\mu$ is likely to be more sensitive to the length, $L$, of the major axis of the rods. 
Interestingly, this approach has revealed to be reliable in identifying those regimes for which the sensitivity of $\mu$ on $L$ is maximized. All in all our results show that $\mu$ can be tuned by properly combining the channel geometry and the length and charge of the rod. 
Moreover, our model allows to rationalize the appearance of non-monotonous dependence of $\mu$ on both, the charge of the rod and its length. Our results can be useful for the design of novel micro- and nano-fluidic devices aiming at sorting stiff filaments. 
%In particular we stress the non-monotonous dependence of $\mu$ on the inverse Debye length $\kappa$. 

\section*{Acknowledgements}
This work was supported by the Deutsche Forschungsgemeinschaft (DFG, German Research Foundation) – Project-ID 416229255 – SFB 1411.

\bibliography{bib_FJ_rods}

\end{document}